\documentclass[10pt,twocolumn,twoside] {IEEEtran}
\ifCLASSINFOpdf
\else
   \usepackage[dvips]{graphicx}
\fi
\IEEEoverridecommandlockouts
\usepackage{algorithm}
\usepackage{graphicx}
\usepackage[noend]{algpseudocode}
\usepackage{url}
\usepackage{breqn,bbm}
\usepackage{cite}
\usepackage{parskip}
\usepackage[usenames]{color}
\usepackage{amsfonts}
\usepackage{fancyhdr}
\usepackage{todonotes,comment}
\usepackage{arydshln,mathtools}
\usepackage{amsmath,amssymb,amsthm}
\usepackage{psfrag,caption,subcaption}
\usepackage{blkarray}
\newtheorem{theorem}{Theorem}

\newtheorem{lemma}{Lemma}

\usepackage{arydshln}
\def\BibTeX{{\rm B\kern-.05em{\sc i\kern-.025em b}\kern-.08em
    T\kern-.1667em\lower.7ex\hbox{E}\kern-.125emX}}

\renewcommand{\thefootnote}{\fnsymbol{footnote}}

\def\cast{{
   \mathord{
      \hbox to 0em{
         \ooalign{
	   \smash{\hbox{$\ast$}}\crcr
	   \smash{\hskip-1pt\Large\hbox{$\circ$}} }
	 \hidewidth}
      \phantom{\bigcirc}
} }}

\def\bm#1{\mbox{\boldmath $#1$}}

\newcommand{\bds}{\begin {itemize}}
\newcommand{\eds}{\end {itemize}}
\newcommand{\bdf}{\begin{definition}}
\newcommand{\blm}{\begin{lemma}}
\newcommand{\edf}{\end{definition}}
\newcommand{\elm}{\end{lemma}}
\newcommand{\bthm}{\begin{theorem}}
\newcommand{\ethm}{\end{theorem}}
\newcommand{\bprp}{\begin{prop}}
\newcommand{\eprp}{\end{prop}}
\newcommand{\bcl}{\begin{claim}}
\newcommand{\ecl}{\end{claim}}
\newcommand{\bcr}{\begin{coro}}
\newcommand{\ecr}{\end{coro}}
\newcommand{\bquest}{\begin{question}}
\newcommand{\equest}{\end{question}}

\newcommand{\larrow}{{\larrow}}

\newcommand{\argmin}{\ensuremath{\mathrm{arg}\min}}
\newcommand{\argmax}{\ensuremath{\mathrm{arg}\max}}

\newcommand{\cH}{{\ensuremath{\mathcal{H}}}}

\newcommand{\cN}{{\ensuremath{\mathcal{N}}}}

\newcommand{\cR}{{\ensuremath{\mathcal{R}}}}
\newcommand{\cS}{{\ensuremath{\mathcal{S}}}}

\newcommand{\cW}{{\ensuremath{\mathcal{W}}}}

\newcommand{\vb}{{\ensuremath{{\mathbf{b}}}}}

\newcommand{\vq}{{\ensuremath{{\mathbf{q}}}}}
\newcommand{\vr}{{\ensuremath{{\mathbf{r}}}}}

\newcommand{\vv}{{\ensuremath{{\mathbf{v}}}}}

\newcommand{\vx}{{\ensuremath{{\mathbf{x}}}}}

\newcommand{\vy}{{\ensuremath{{\mathbf{y}}}}}
\newcommand{\vz}{{\ensuremath{{\mathbf{z}}}}}

\newcommand{\mA}{{\ensuremath{\mathbf{A}}}}

\newcommand{\mB}{{\ensuremath{\mathbf{B}}}}

\newcommand{\mC}{{\ensuremath{\mathbf{C}}}}

\newcommand{\mL}{{\ensuremath{\mathbf{L}}}}

\newcommand{\mQ}{{\ensuremath{\mathbf{Q}}}}

\newcommand{\mU}{{\ensuremath{\mathbf{U}}}}

\newcommand{\mV}{{\ensuremath{\mathbf{V}}}}

\newcommand{\mY}{{\ensuremath{\mathbf{Y}}}}
\newcommand{\mZ}{{\ensuremath{\mathbf{Z}}}}

\usepackage{latexsym}

\def\IC{\mathbb C}
\def\IN{\mathbb N}
\def\IZ{\mathbb Z}
\def\IR{\mathbb R}

\def\shat{^{\mathchoice{}{}%
 {\,\,\smash{\hbox{\lower4pt\hbox{$\widehat{\null}$}}}}%
 {\,\smash{\hbox{\lower3pt\hbox{$\hat{\null}$}}}}}}

\def\bSigma{{
      \ooalign{
      \smash{\hskip.4pt\raise.4pt\hbox{$\Sigma$}}\vphantom{}\crcr
      \smash{\hskip.7pt\raise.6pt\hbox{$\Sigma$}}\vphantom{}\crcr
      \smash{\hbox{$\Sigma$}}\vphantom{$\Sigma$}}
      \vphantom{\hbox{$\Sigma$}}
      }}
\def\bTheta{{
      \ooalign{
      \smash{\hskip.5pt\raise.5pt\hbox{$\Theta$}}\vphantom{}\crcr
      \smash{\hskip.0pt\raise.1pt\hbox{$\Theta$}}\vphantom{}\crcr
      \smash{\hbox{$\Theta$}}\vphantom{$\Theta$}}
      \vphantom{\hbox{$\Theta$}}
      }}
\def\bDelta{{
      \ooalign{
      \smash{\hskip.4pt\raise.4pt\hbox{$\Delta$}}\vphantom{}\crcr
      \smash{\hskip.7pt\raise.6pt\hbox{$\Delta$}}\vphantom{}\crcr
      \smash{\hbox{$\Delta$}}\vphantom{$\Delta$}}
      \vphantom{\hbox{$\Delta$}}
      }}
\def\bLambda{{
      \ooalign{
      \smash{\hskip.5pt\raise.5pt\hbox{$\Lambda$}}\vphantom{}\crcr
      \smash{\hskip.0pt\raise.1pt\hbox{$\Lambda$}}\vphantom{}\crcr
      \smash{\hbox{$\Lambda$}}\vphantom{$\Lambda$}}
      \vphantom{\hbox{$\Lambda$}}
      }}

\makeatletter

\def\bordermatrix#1{\begingroup \m@th
  \@tempdima 8.75\p@
  \setbox\z@\vbox{%
    \def\cr{\crcr\noalign{\kern2\p@\global\let\cr\endline}}%
    \ialign{$##$\hfil\kern2\p@\kern\@tempdima&\thinspace\hfil$##$\hfil
      &&\quad\hfil$##$\hfil\crcr
      \omit\strut\hfil\crcr\noalign{\kern-\baselineskip}%
      #1\crcr\omit\strut\cr}}%
  \setbox\tw@\vbox{\unvcopy\z@\global\setbox\@ne\lastbox}%
  \setbox\tw@\hbox{\unhbox\@ne\unskip\global\setbox\@ne\lastbox}%
  \setbox\tw@\hbox{$\kern\wd\@ne\kern-\@tempdima\left[\kern-\wd\@ne
    \global\setbox\@ne\vbox{\box\@ne\kern2\p@}%
    \vcenter{\kern-\ht\@ne\unvbox\z@\kern-\baselineskip}\,\right]$}%
  \null\;\vbox{\kern\ht\@ne\box\tw@}\endgroup}
\makeatother

\makeatletter
\def\argmin{\mathop{\operator@font arg\,min}}
\def\argmax{\mathop{\operator@font arg\,max}}
\makeatother

\newcommand{\diag}{\mbox{\rm diag}}

\def\bm#1{\mbox{\boldmath $#1$}}

\newcommand{\bbeta}{{\bm\beta}}

\newcommand{\bea}{\begin{array}}
\newcommand{\ena}{\end{array}}
\newcommand{\beq}{\begin{equation}}
\newcommand{\enq}{\end{equation}}

\newcommand{\beqa}{\begin{eqnarray}}
\newcommand{\enqa}{\end{eqnarray}}

\newcommand{\beqan}{\begin{eqnarray*}}
\newcommand{\enqan}{\end{eqnarray*}}

\newcommand{\AL}{\begin{enumerate}}
\newcommand{\ALE}{\end{enumerate}}

\def\addots{\mathinner{
    \mkern1mu\raise0pt\vbox{\kern7pt\hbox{.}}
    \mkern2mu\raise4pt\hbox{.}
    \mkern2mu\raise7pt\hbox{.}
    \mkern1mu}}

\def\sddots{\mathinner{
    \mkern.8mu\raise7pt\hbox{.}
    \mkern.8mu\raise4pt\hbox{.}
    \mkern.8mu\raise0pt\vbox{\kern7pt\hbox{.}}
    \mkern1mu}}

\def\saddots{\mathinner{
    \mkern.2mu\raise0pt\vbox{\kern7pt\hbox{.}}
    \mkern.2mu\raise4pt\hbox{.}
    \mkern.2mu\raise7pt\hbox{.}
    \mkern1mu}}

\def\sqplus{\mathbin{
	{\ooalign{\hfil\raise.3ex\hbox{\scriptsize
	+}\hfil\crcr\mathhexbox274\crcr\mathhexbox275}}
	}} 
\def\sqminus{\mathbin{
	{\ooalign{\hfil\raise.3ex\hbox{\scriptsize
	--}\hfil\crcr\mathhexbox274\crcr\mathhexbox275}}
	}}

\def\IC{{
   \mathord{
      \hbox to 0em{
	 \hskip-4pt
         \ooalign{
	   \smash{\hskip1.9pt\raise2.6pt\hbox{$\scriptscriptstyle |$}}\crcr
	   \smash{\hbox{\rm\sf C}} }
	 \hidewidth}
      \phantom{\hbox{\rm\sf C}}
} }}
\def\IN{
    {\ooalign{
   \smash{\hskip2.2pt\raise1.5pt\hbox{$\scriptscriptstyle |$}}\vphantom{}\crcr
   \hbox{\sf N}
	}}
	} 
\def\IZ{
    {\ooalign{
   \smash{\hskip1.9pt\raise0pt\hbox{$\sf Z$}}\vphantom{}\crcr
   \hbox{\sf Z}
	}}
	} 
\def\IR{
    {\ooalign{
   \smash{\hskip2.2pt\raise1.5pt\hbox{$\scriptscriptstyle |$}}\vphantom{}\crcr
   \smash{\hskip2.2pt\raise3.3pt\hbox{$\scriptscriptstyle |$}}\vphantom{}\crcr
   \hbox{\sf R}
	}}
	} 

\DeclareMathAlphabet{\mathcmb}{OT1}{cmr}{b}{n}

\def\bSigma{\ensuremath{\mathcmb{\Sigma}}}
\def\bLambda{\ensuremath{\mathcmb{\Lambda}}}

\def\bTheta{\ensuremath{\mathcmb{\Theta}}}

\newcommand{\SI}{\begin{indlist}}
\newcommand{\EI}{\end{indlist}}

\newcommand{\DL}{\begin{dashlist}}
\newcommand{\DLE}{\end{dashlist}}

\makeatletter
\def\setboxz@h{\setbox\z@\hbox}
\def\wdz@{\wd\z@}
\def\boxz@{\box\z@}
\def\underset#1#2{\binrel@{#2}%
  \binrel@@{\mathop{\kern\z@#2}\limits_{#1}}}
\def\binrel@#1{\begingroup
  \setboxz@h{\thinmuskip0mu
    \medmuskip\m@ne mu\thickmuskip\@ne mu
    \setbox\tw@\hbox{$#1\m@th$}\kern-\wd\tw@
    ${}#1{}\m@th$}%
  \edef\@tempa{\endgroup\let\noexpand\binrel@@
    \ifdim\wdz@<\z@ \mathbin
    \else\ifdim\wdz@>\z@ \mathrel
    \else \relax\fi\fi}%
  \@tempa
}
\let\binrel@@\relax%
\makeatother

\begin{document}

\title{Sampling and Recovery of Signals on a Simplicial Complex using Neighbourhood Aggregation}

\author{Siddartha Reddy and Sundeep Prabhakar Chepuri, \IEEEmembership{Member, IEEE}
\thanks{S. Reddy and S.P. Chepuri are with the Department of ECE, Indian Institute of Science, Bengaluru 560012, India (e-mail: \{thummalurur,spchepuri\}@iisc.ac.in).}}

\maketitle
\thispagestyle{empty}
\pagenumbering{arabic}
\renewcommand{\thefootnote}{\arabic{footnote}}
\IEEEoverridecommandlockouts
\maketitle
\begin{abstract}
In this work, we focus on sampling and recovery of signals over simplicial complexes.  In particular, we subsample a simplicial signal of a certain order and focus on recovering multi-order bandlimited simplicial signals of one order higher and one order lower. To do so, we assume that the simplicial signal admits the Helmholtz decomposition that relates simplicial signals of these different orders. Next, we propose an aggregation sampling scheme for simplicial signals based on the Hodge Laplacian matrix and a simple least squares estimator for recovery. We also provide theoretical conditions on the number of aggregations and size of the sampling set required for faithful reconstruction as a function of the bandwidth of simplicial signals to be recovered.  Numerical experiments are provided to show the effectiveness of the proposed method.

\end{abstract}

\section{Introduction} \label{sec:introduction}

\IEEEPARstart{S}{implicial complexes} are structures that capture topological information by modeling higher-order (beyond pairwise)  interactions in data. Graphs may be viewed as a specialization of simplicial complexes that capture only pairwise interactions between entities (as nodes) through edges. A simplicial complex is a finite collection of simplicies, where a $k$-simplex (or a simplex of order $k$) is a subset of a vertex set with cardinality $k+1$. For example, a node is a $0$-simplex, an edge is a $1$-simplex, a closed triangle is a 2-simplex, and so on. The neighborhood information of simplicies of different orders in a simplicial complex is captured by the (higher-order) Hodge Laplacian matrix.  While signals enumerated by the nodes of a graph are called graph signals, signals indexed using different simplices in a simplicial complex are referred to as simplicial signals. For instance, in contact networks~\cite{benson2018simplicial},   interactions between individuals are modeled using simplicial complexes with nodes being the individuals and signals being the total number of times a group of two or more individuals came in contact with each other. Generalization of graph signal processing~\cite{ortega2018graph,GSP} tools to simplicial complex data such as filtering, learning, and sampling is gaining attention~\cite{TSP,schaub2020random,reddy2023clustering,jia2019graph} for problems like trajectory prediction, simplicial closure, clustering, and edge flow denoising, to name a few. 

Sampling and recovery of signals enumerated by a graph (\emph{graph sampling}, in short) extends time-domain sampling to irregular (non-Euclidean) domains. Graph sampling is a well-studied problem~\cite{tanaka2020sampling,tsitsvero2016signals,reddy2019sampling,chepuri2017graph,sakiyama2019eigendecomposition,anis2016efficient,segarra2016reconstruction}, where the key idea is to recover graph signals on all the nodes by observing only a subset of them. To do so, graph signals are assumed to be bandlimited or smooth, which, in other words, means that the graph signals can be synthesized using a few eigenmodes of the Laplacian matrix of the underlying graph. A closely related graph sampling mechanism, which we extend to simplicial signals in this work is the aggregation sampling method~\cite{marques2015sampling}, wherein the algorithm recovers graph signals on all the nodes by observing information aggregated (from different neighborhoods using a graph aggregation operator) at a few nodes. 

In this work, we focus on the problem of sampling and recovery of simplicial signals~\cite{ebli2020simplicial} through a neighborhood aggregation mechanism that gathers information from upper and lower adjacent neighbors (in contrast, for nodal signals, there are no lower adjacent neighbors). Specifically, we restrict our attention to discretized smooth vector fields or bandlimited edge flow signals that admit the Helmholtz decomposition. Leveraging the Helmholtz decomposition while aggregating and sampling enables recovery of multi-order simplicial signals, i.e., simplicial signals of one order lower (say, order $k-1$) and one order higher (say, order $k+1$) by observing a few aggregated $k$-simplicial signals. Existing works~\cite{TSP,jia2019graph} on sampling and recovery of simplicial signals focus on  
sampling and recovery of edge flows signals or recovery of single-order simplicial signals and do not consider simultaneous multi-order simplicial signal recovery. In particular,~\cite{jia2019graph} extends the idea of node signal recovery to recover the signals on the edges (i.e., $1$-simplex), and~\cite {TSP}, on the other hand, proposes an algorithm to recover $1$-simplicial signals by sampling multi-order simplicial signals leveraging the relationship among the signals using the Helmholtz decomposition. To the best of our knowledge, this is the first work that considers recovery of multi-order simplicial signals from subsamples of a single-order simplicial signal.

The main contributions in the letter are (a) the proposed sampling scheme based on aggregation of edge flow signals through the Hodge Laplacian matrix, (b) least squares-based recovery of multi-order bandlimited simplicial signals, and (c) theoretical guarantees on the number of aggregations and the size of the sampling set as a function of the bandwidth of multi-order signals for perfect recovery. 


\noindent \textbf{Notation}:
We use the following notation throughout the paper. We use upper (lower) boldface letters for matrices (column vectors) and calligraphic letters to denote sets. We denote matrix vectorization operation by ${\rm vec}(\cdot)$, transpose of a matrix by   $(\cdot)^{T}$, range space of a matrix by $\mathcal{R}(\cdot)$, null space of a matrix by $\mathcal{N}(\cdot)$, direct sum by $\oplus$, and Khatri-Rao product (i.e., a column-wise Kronecker product) by $\odot$.  We frequently use the identity $\rm{vec}(\mA\rm{diag}(\vb)\mC) = \left(\mC^{T} \odot \mA\right)\vb$.

\section{Signals over Simplicial Complexes} \label{sec:preliminaries}
Consider simplicial complexes that contain up to 2-simplices whose topological structure is described by the so-called \emph{Hodge (edge) Laplacian matrix}:
\begin{align}
\mL_{1} = \underbrace{\mB_{1}^T\mB_{1}}_{:=\mL_{\rm low}} + \underbrace{\mB_{2}\mB_{2}^T}_{:=\mL_{\rm up}} \in \mathbb{R}^{N_1 \times N_1},
\label{eq:HodgeLaplacian}
\end{align}
where $\mB_k \in \mathbb{R}^{N_{k-1}\times N_{k}}$ is the incidence matrix. The node and triangle Laplacian matrices are given as 
\[ \mL_{0} = \mB_{1}\mB_{1}^{T}\in \mathbb{R}^{N_0 \times N_0} \quad \text{and} \quad \mL_{2} = \mB_{2}^{T}\mB_{2} \in \mathbb{R}^{N_2 \times N_2}, 
\]
respectively. These Laplacian matrices admit the spectral decomposition $\mL_k = \mQ_k{\boldsymbol \Lambda}_k\mQ_k^T$ for $k=0,1,2$, where $\mQ_k$ collects the eigenvectors and the diagonal matrix $\boldsymbol{\Lambda}_{k} = \diag(\lambda_{1,k},\cdots,\lambda_{N_k,k})$ collects the corresponding eigenvalues of $\mL_k$. We also define $\mL_{\rm low} = \mU_{\rm low}{\boldsymbol \Lambda}_{\rm low}\mU_{\rm low}^T$ and $\mL_{\rm up} = \mU_{\rm up}{\boldsymbol \Lambda}_{\rm up}\mU_{\rm up}^T$, where $\mU_{\rm low}$ and ${\boldsymbol \Lambda}_{\rm low}$ (resp.,  $\mU_{\rm up}$ and ${\boldsymbol \Lambda}_{\rm up}$) collect the eigenvectors and eigenvalues of $\mL_{\rm low}$ (resp., $\mL_{\rm up}$). The edge Laplacian matrix decomposes the $N_1$-dimensional space into three orthogonal subspaces as $\mathbb{R}^{N_{1}}= \mathcal{R}(\mB^{T}_{1}) \, \oplus\, \mathcal{R}(\mB_{2}) \, \oplus\, \mathcal{N}(\mL_{1})$. Thus we have $\mB_{1} \mB_{2} = \mathbf{0}$ or $\mL_{\rm low}\mL_{\rm up} = \mathbf{0}$.
 
A $k$-simplicial signal, denoted by $\vx_k \in \mathbb{R}^{N_k}$, is indexed by a $k$-simplex, e.g., a $0$-simplicial signal is a node signal and a $1$-simplicial signal is an edge flow (or a discretized vector field) signal. We assume that $k$-simplicial signals admit the Helmholtz decomposition, which allows us to express 1-simplicial signals in terms of $0$- and $2$- simplicial signals with a  residual component (also a $1$-simplicial signal) as 
\begin{equation}
    \vx_{1} = \mB^{T}_{1}\vx_{0}+\mB_{2}\vx_{2}+\vr_{1} \in \mathbb{R}^{N_1},
    \label{eq:Hodge_sig_decomp}
\end{equation}
where $\vr_{1} \in \mathbb{R}^{N_1}$ satisfies~$\mL_{1}\vr_{1}=~{\bf 0}$.  We can write $\vx_{k} = \mQ_k\underline{\hat{\vx}}_{k}$ for $k=0,1,2$ and $\vr_1 = \mQ_1\underline{\hat{\vr}}_{1}$, where 
$\underline{\hat{\vx}}_{k}$ and $\underline{\hat{\vr}}_{1}$ are the simplicial spectral representation of $\vx_k$ and $\vr_1$, respectively.

The simplicial signal $\vx_{k}$ is said to be $W_k$\emph{-bandlimited} if it satisfies \[\vx_{k} = \tilde{\mQ}_k \hat{\vx}_{k}\] for $k=0,1,2$, where 
$\tilde{\mQ}_{k}$ collects a subset of columns (w.l.o.g., say the first $W_k$ columns) of $\mQ_k$ corresponding to $\cR(\mL_k)$ and 
$\underline{\hat{\vx}}_{k} = [\hat{\vx}_{k}^T, {\bf 0}^T]^T$ with $\hat{\vx}_{k} \in \mathbb{R}^{W_k}$. One can show that (cf. Appendix or~\cite{TSP})
\begin{align}
\tilde{\mU}_{\rm low} &= \mB_{1}^{T}\tilde{\mQ}_{0} \in \mathbb{R}^{N_1 \times W_0}, \notag\\
\quad \tilde{\mU}_{\rm up} &= \mB_{2}\tilde{\mQ}_{2}\in \mathbb{R}^{N_1 \times W_2},
\label{eq:eigenvector_relations}
\end{align}
which are eigenvectors of $\mL_{\rm low}$ and $\mL_{\rm up}$, associated with $\cR(\mL_1)$, expressed using the eigenvectors of $\mL_0$ and $\mL_2$, respectively.
Thus $\mL_{\rm low}\tilde{\mU}_{\rm low} = \tilde{\mU}_{\rm low}\tilde{\boldsymbol{\Lambda}}_{\rm low}$ and $\mL_{\rm up}\tilde{\mU}_{\rm low} = \tilde{\mU}_{\rm up}\tilde{\boldsymbol{\Lambda}}_{\rm up}$, where $\tilde{\boldsymbol{\Lambda}}_{\rm low} = \diag(\lambda_{1,{\rm low}}, \cdots,\lambda_{W_0,{\rm low}})$ and $\tilde{\boldsymbol{\Lambda}}_{\rm up} = \diag(\lambda_{1,{\rm up}}, \cdots,\lambda_{W_2,{\rm up}})$ contain non-zero eigenvalues of $\mL_{\rm low}$ and $\mL_{\rm up}$, respectively. 

Assuming that $\vx_{1}$ admits the decomposition in~\eqref{eq:Hodge_sig_decomp} and can be synthesized using $W_{0}$-bandlimited $\vx_{0} \in \cR(\mL_{0})$, $W_{2}$-bandlimited
 $\vx_{2} \in \cR(\mL_{2})$, we have $W_1 = W_0+W_2 +R_1$, where the remaining basis vectors corresponding to the $\cN(\mL_1)$ restrict the bandwidth of $\vr_1$ to  $R_1$ with 
\begin{equation}
\vr_1 = \tilde{\mQ}_1^{\perp}\hat{\vr}_1.
\end{equation}
Here, $\tilde{\mQ}_1^{\perp}$ collects a subset of columns of 
$\mQ_1$ corresponding to $\cN(\mL_1)$ and $\hat{\vr}_1 \in \mathbb{R}^{R_1}$.
\section{Aggregation of Simplicial Signals} \label{sec:aggregation}

In this section, we discuss the proposed approach to reconstruct multi-order simplicial signals. We also provide conditions for perfect recovery. 

\subsection{The Sampling Problem}
Consider the following model for aggregating the $1$-simplicial signal $\vx_1$ via the $p$-th integer powers of the edge Laplacian matrix $\mL_1$ to obtain the $p$-th shifted $1$-simplicial signal as
\begin{align}
\vy^{(p)}_{1} &= \mL^{p}_{1}\vx_{1}  \overset{(a)}{=} (\mL_{\rm low} + \mL_{\rm up})^{p} \vx_{1}
\label{eq:simplicial_shifting}
\end{align}
with $\vy^{(0)}_{1} = \vx_{1}$, where we can see above in $(a)$ that the shifted edge signal is an aggregation from the lower-order and upper-order simplicial neighbors of the $1$-simplex. Substituting the Helmholtz  decomposition~\eqref{eq:Hodge_sig_decomp}, we arrive at a model that relates \emph{multi-order simplicial signals} as
\begin{align}
\vy^{(p)}_{1} &= \mL_{\rm low}^{p}\mB^{T}_{1}\vx_{0}+\mL_{\rm up}^{p}\mB_{2}\vx_{2} +\mL_{1}^{p}\vr_{1},
\label{eq:simplicial_shifting2}
\end{align}
which satisfies~\eqref{eq:Hodge_sig_decomp} when $p=0$. To arrive at the above simplified equation, we use the fact that $\mL_{\rm low}\mL_{\rm up} ={\bf 0}$. We collect the $1$-simplicial shifted signals for $P$ shifts to obtain  
the $N_1 \times P$ matrix 
\begin{equation}
  \mY_1 = [\vy^{(0)}_{1}, \vy^{(1)}_{1}, \cdots,\vy^{(P-1)}_{1}],
\end{equation}
where $P$ denotes the number of diffusions/aggregations (i.e., the information gathered from $P$-hop neighborhood). 

We only observe a few edge flows using the sampling matrix $\boldsymbol{\Phi}$, which selects rows of $\mY_1$ indexed by the sampling set $\cS$, i.e., we observe
\[
\mZ_{1}= \boldsymbol{\Phi}\mY_{1} \, \in \mathbb{R}^{|\cS| \times P},
\]
where $\boldsymbol{\Phi} \in \{0,1\}^{|\cS| \times N_{1}}$. This means that when $|\cS|= 1$, we observe only one aggregated $1$-simplicial signal (or one aggregated edge flow) at one of the edges. Given the subsampled observations $\mZ_{1}$ and the sampling matrix $\boldsymbol{\Phi}$, we are interested in recovering bandlimited simplicial signals $\vx_{0}$, $\vx_{2}$, and $\vr_{1}$.

\subsection{Recovery of Multi-order Simplicial Signals}\label{sec:multilayer}

Using the spectral representation $\vx_{k} = \tilde{\mQ}_{k}\hat{\vx}_{k}$ and $\vr_{1} = \tilde{\mQ}_{1}^\perp\hat{\vr}_{1}$, we can re-write~\eqref{eq:simplicial_shifting2} to relate $\vy_1^{(p)}$ to the spectral domain of the underlying simplicial signals as
\begin{align}
\vy^{(p)}_{1}
&=  
\underbrace{\left[ 
\begin{array}{lll}
\mL_{\rm low}^{p}\mB_{1}^{T}\tilde{\mQ}_{0} & \mL_{\rm up}^{p} \tilde{\mQ}_{2}\mB_{2} & \mL_{1}^{p}\tilde{\mQ}_{1}^\perp
\end{array}
\right]}_{:=\mA^{(p)}}
\underbrace{\left[
\begin{array}{l}
    \hat{\vx}_{0}\\
   \hat{\vx}_{2}\\
    \hat{\vr}_{1}
\end{array}
\right]}_{:=\hat{\vx}}. \notag
\end{align}
From~\eqref{eq:eigenvector_relations}, we can further simplify $\mA^{(p)}$ as
\begin{align}
\mA^{(p)}
&=  \left[
\begin{array}{lll}
\tilde{\mU}_{\rm low} \tilde{\bLambda}_{\rm low}^{p}& \tilde{\mU}_{\rm up}\tilde{\bLambda}_{\rm up}^{p} &\mL_{1}^{p}\tilde{\mQ}_{1}^\perp
\end{array}
\right] \notag\\
&= \left[
\begin{array}{lll}
\tilde{\mU}_{\rm low} & \tilde{\mU}_{\rm up} & \tilde{\mQ}_{1}^\perp
\end{array}
\right]
\diag({\vv}^{(p)}) \notag
\end{align}
where recall that the eigenvector matrix $\tilde{\mQ}_{1}^\perp$ is associated to the zero eigenvalues of $\mL_1$ with $\mL_{1}^{p}\tilde{\mQ}_{1}^\perp = \tilde{\mQ}_{1}^\perp$ when $p=0$ and $\mL_{1}^{p}\tilde{\mQ}_{1}^\perp = {\bf 0}$, otherwise. Here, we have introduced $\vv^{(p)} = [\lambda_{1,{\rm low}}^p, \cdots,\lambda_{W_0,{\rm low}}^p, \lambda_{1,{\rm up}}^p, \cdots,\lambda_{W_2,{\rm up}}^p, \mathbbm{1}_p]^T$, where the indicator vector $\mathbbm{1}_p$ is the all-one vector when $p=0$ and is all-zero, otherwise. Thus we have
\begin{align}
\vy^{(p)}_{1} = 
\left[
\begin{array}{lll}
\tilde{\mU}_{\rm low} & \tilde{\mU}_{\rm up} & \tilde{\mQ}_{1}^\perp
\end{array}
\right] \diag(\hat{\vx})\,{\vv}^{(p)}, \notag
\end{align}
which for $p=0,1,\ldots,P-1$ yields
\begin{align}
\mZ_{1} =  \boldsymbol{\Phi}
\left[
\begin{array}{lll}
\tilde{\mU}_{\rm low} & \tilde{\mU}_{\rm up} & \tilde{\mQ}_{1}^\perp
\end{array}
\right] \diag(\hat{\vx})\,\mV, 
\label{eq:final_Z1}
\end{align}
where $\mV$ is the Vandermonde matrix formed with  $\vv^{(p)}$ as
\[
\mV = 
\left[\begin{array}{cccc}
1 & \lambda_{1,{\rm low}} & \cdots & \lambda_{1,{\rm low}}^{P-1} \\
\vdots & \vdots & \cdots & \vdots \\
1 & \lambda_{W_0,{\rm low}} & \cdots & \lambda_{W_0,{\rm low}}^{P-1} \\ \hdashline
1 & \lambda_{1,{\rm up}} & \cdots & \lambda_{1,{\rm up}}^{P-1} \\
\vdots & \vdots & \cdots & \vdots \\1 & \lambda_{W_2,{\rm up}} & \cdots & \lambda_{W_2,{\rm up}}^{P-1} \\ 
\hdashline {\bf 1} & {\bf 0} & \cdots & {\bf 0}
\end{array}
\right] \in \mathbb{R}^{W_1 \times P}.
\]

On vectorizing~\eqref{eq:final_Z1}, we get a linear system having $P|\cS|$ equations in $W_1$ unknowns
\[
\vz_1 = \left(\mV^T \odot {\boldsymbol \Phi} [\tilde{\mU}_{\rm low},\,\, \tilde{\mU}_{\rm up} ,\,\, \tilde{\mQ}_{1}^\perp]\right) \hat{\vx}.
\]
If the matrix $\mV^{T}\odot \boldsymbol{\Phi}[\tilde{\mU}_{\rm low},\,\,\tilde{\mU}_{\rm up},\,\,\tilde{\mQ}_1^\perp] \in \mathbb{R}^{P|\cS| \times W_{1}}$ has  full column rank, then the above system can be solved uniquely using least squares as  
\begin{align}
\begin{bmatrix}
\hat{\vx}_{0,{\rm ls}}\\
\hat{\vx}_{2,{\rm ls}}\\
\hat{\vr}_{1,,{\rm ls}}\\
\end{bmatrix} = \left(\mV^{T} \odot \boldsymbol{\Phi}\,[\tilde{\mU}_{\rm low},\,\,\tilde{\mU}_{\rm up},\,\,\tilde{\mQ}_1^\perp]\right)^{\dagger}\vz_{1}.
\label{eq:mlayer8}
\end{align}
Finally, the simplicial signals can be recovered as
\begin{align}
    {\vx}_{0,{\rm ls}} = \tilde{\mQ}_{0}\hat{\vx}_{0,{\rm ls}},\,
    {\vx}_{2,{\rm ls}} = \tilde{\mQ}_{2}\hat{\vx}_{2,{\rm ls}},\, \text{and} \,\,
    {\vr}_{1,{\rm ls}} = \tilde{\mQ}_1^\perp\hat{\vr}_{1,{\rm ls}}.
     \notag
\end{align}
Next, we provide conditions on $P$ and $|\cS|$ for perfect recovery of multi-order simplicial signals as in the following theorem.

\begin{theorem}\label{theorem} Suppose $\vx_{1}$ is $W_{1}$-bandlimited and the eigenvalues $\{\lambda_{1,{\rm low}}, \cdots,$ $\lambda_{W_0,{\rm low}},\lambda_{1,{\rm up}}, \cdots,\lambda_{W_2,{\rm up}}\}$ are distinct. Assume that $\vx_0$, $\vx_2$ and $\vr_1$ are $W_0$-, $W_2$-, and $R_1$-bandlimited, respectively. Then $\vx_{0}, \vx_{2}$, and $\vr_{1}$ can be perfectly reconstructed from the subsampled observations $\vz_{1}$ obtained with at least $|\mathcal{S}| = R_{1}$ and $P =  W_{0}+ W_{2}+1$. 
\end{theorem}
\begin{figure*}[t]
\centering
\includegraphics[width=1.6\columnwidth]{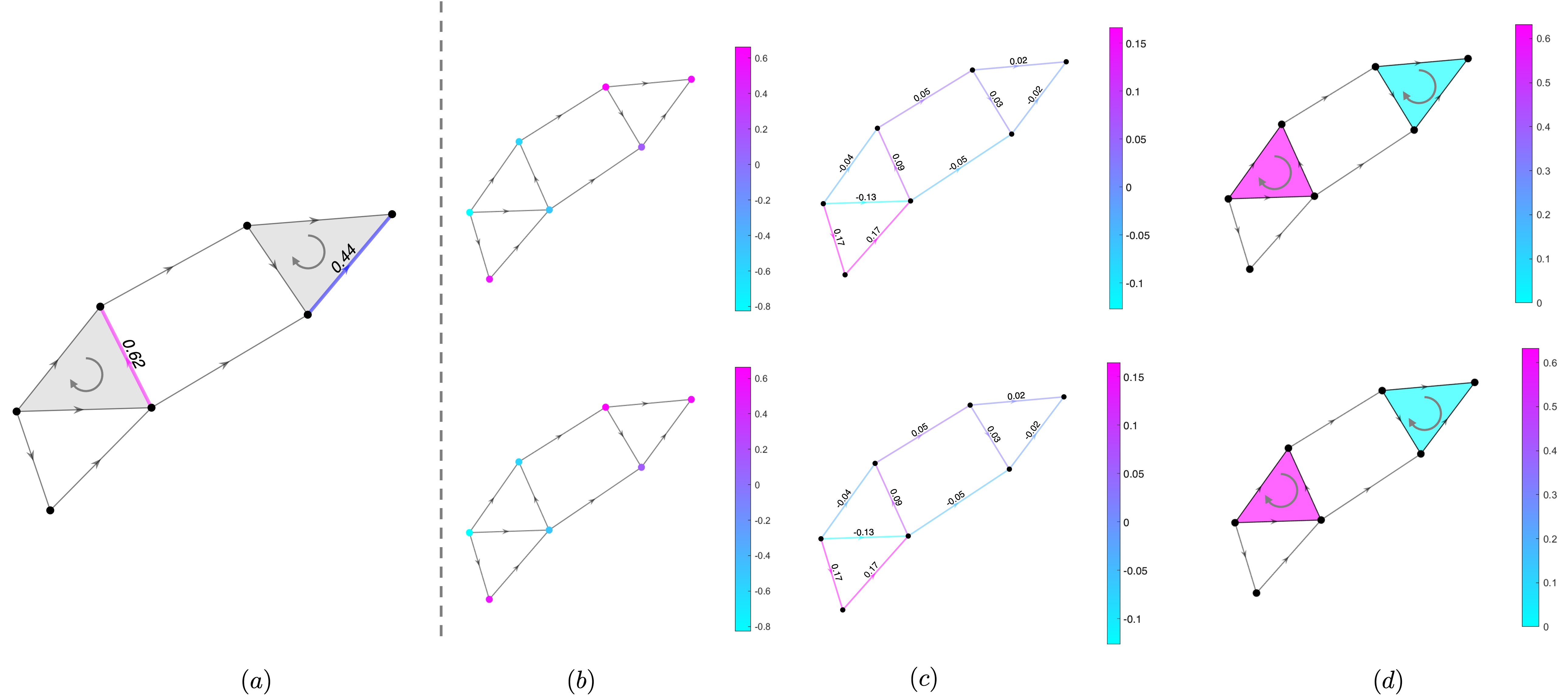}
\caption{\small Synthetic dataset. (a) Observed edge flows indicated by the two colored arrows with the flow values on the top. (b) Bandlimited node signal $\vx_0$: unobserved true node signal (top) and recovered node signal (bottom). (c) Bandlimited residual signals $\vr_{1}$: unobserved true residual signal (top) and recovered residual signal (bottom). (d) Bandlimited triangle signal $\vx_2$: unobserved true triangle signal (top) and recovered triangle signal (bottom).}
\label{fig:fig1}\hfill
\end{figure*} 
\begin{figure*}
\centering
\includegraphics[width=1.75\columnwidth]{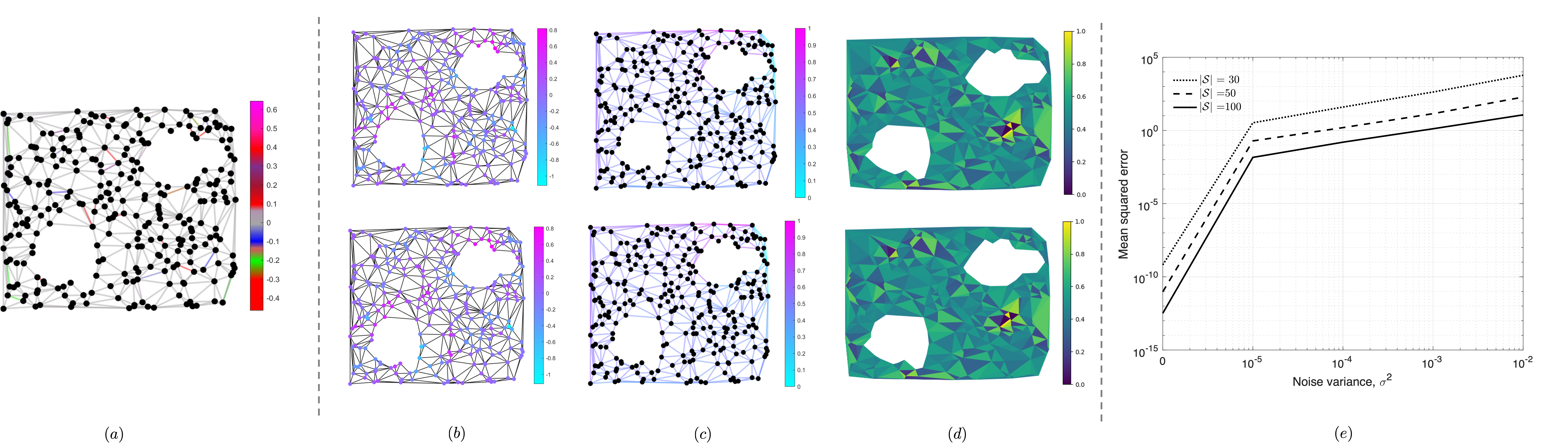}
\caption{Two-hole dataset.  (a) Observed edge flow. (b Bandlimited node signal $\vx_0$: unobserved true node signal (top) and recovered node signal (bottom). (c) Bandlimited residual signal $\vr_{1}$: unobserved true residual signal (top) and recovered residual signal (bottom). (d) Bandlimited triangle signal $\vx_2$: unobserved true triangle signal (top) and recovered triangle signal (bottom). (e) Mean squared error for different values of noise variance and different values of $|\cS|$ with $P=10$.}
\label{fig:fig2}
\end{figure*} 

\begin{proof}  The matrix $\mV^{T}\odot \boldsymbol{\Phi}[\tilde{\mU}_{\rm low},\,\,\tilde{\mU}_{\rm up},\,\,\tilde{\mQ}_1^\perp]$ has full column rank, if $k\text{rank}(\mV^{T}) + {k}\text{rank}(\boldsymbol{\Phi}[\tilde{\mU}_{\rm low},\,\,\tilde{\mU}_{\rm up},\,\,\tilde{\mQ}_1^\perp]) \geq W_1+1$~\cite{sidiropoulos2000parallel}, where ${k}\text{rank}$ is the Kruskal rank. When $P>W_1$ and the eigenvalues in $\{\lambda_{1,{\rm low}}, \cdots,$ $\lambda_{W_0,{\rm low}},\lambda_{1,{\rm up}}, \cdots,\lambda_{W_2,{\rm up}}\}$ are distinct, then the rank of $\mV$ is $W_0+W_2+1$. Since $k\text{rank}(\mV^{T}) \leq \rm{rank}(\mV^{T})$, we require  ${k}\text{rank}(\boldsymbol{\Phi}[\tilde{\mU}_{\rm low},\,\,\tilde{\mU}_{\rm up},\,\,\tilde{\mQ}_1^\perp]) \geq R_1.$
Since the eigenvalues are distinct, the matrix $[\tilde{\mU}_{\rm low},\,\,\tilde{\mU}_{\rm up},\,\,\tilde{\mQ}_1^\perp]$ contains linearly independent vectors and is full column rank. Thus, for $\text{rank}(\boldsymbol{\Phi}[\tilde{\mU}_{\rm low},\,\,\tilde{\mU}_{\rm up},\,\,\tilde{\mQ}_1^\perp]) = R_1$, we need $|\cS|=R_1$ so that $\boldsymbol{\Phi}$ selects $R_1 < W_1$ linearly independent rows of $[\tilde{\mU}_{\rm low},\,\,\tilde{\mU}_{\rm up},\,\,\tilde{\mQ}_1^\perp]$. Hence, for perfect recovery, we need at least $|\cS| = R_1$ and $P = W_{0} + W_{2}+1$. 
\end{proof}

\section{Numerical Experiments} \label{sec:synthetic_Data} 

In this section, we discuss numerical results of the proposed reconstruction method of multi-order bandlimited signals, i.e., recovery of $\{\vx_0,\vx_2,\vr_{1}\}$ from subsamples of aggregated $\vx_1$ on synthetic and two-hole datasets.

To begin with, we consider a simplicial complex that contains up to 2-simplices as shown in  Fig.~\ref{fig:fig1}(a). This simplicial complex has $N_{0} = 7$ nodes, $N_{1} = 10$ edges (orientation indicated with arrows), $N_{2} = 2$ closed triangles (indicated as shaded regions along along with their orientation). For this example, $\mL_0$ has 6 distinct eigenvalues, $\mL_{1}$ has $2$ zero eigenvalues, and $\mL_{2}$ has $1$ distinct eigenvalue.  We generate bandlimited simplicial signals as follows: $W_{0}$-bandlimited signals on nodes, i.e., $\vx_{0}$ are generated as a random linear combination of the $W_{0}$ basis vectors $\mQ_{0}$. Figure~\ref{fig:fig1}(b) (top) shows the bandlimited node signal with bandwidth $W_{0} = 4$. We follow a similar procedure to obtain a $W_{2}$-bandlimited signal  on triangles with $W_{2}=1$, and it is shown in Figure~\ref{fig:fig1}(d) (top).  The $R_{1}$-bandlimited residual signals on the $1$-simplices are obtained by taking a random linear combination of basis vectors of $\cN(\mL_{1})$. We set $R_{1} = 2$ Figure~\ref{fig:fig1}(c) (top) shows bandlimited residual signals on the simplicial complex. After we generate the bandlimited signals $\vx_{0}$, $\vx_{2}$ and $\vr_{1}$,  we compute 
$\vx_1$ and $\mY_{1}$ according to \eqref{eq:Hodge_sig_decomp} and \eqref{eq:simplicial_shifting2}, respectively, for $P=6$ shifts. 
Finally, we obtain subsampled observations by selecting $2$ edges out of $10$ uniformly at random. The size of sampling set and the number of shifts are set as per Theorem~\ref{theorem}, i.e., we use $P= W_0 + W_2 +1$ and $|\cS| = R_{1}$.
The bottom row of Figs.~\ref{fig:fig1}(b),~\ref{fig:fig1}(c), and~\ref{fig:fig1}(d) show the reconstructed node, triangle, and residual signals on simplicial complex, where it is clear that recovery is perfect.  

While the above example was a controlled one, wherein there are distinct eigenvalues that allow us to sample according to the conditions prescribed in Theorem~\ref{theorem}, we next consider a synthetic noisy \emph{two-hole} dataset resembling the ocean drifters phenomenon ~\cite{roddenberry2021principled}, we see that one has to often choose more samples to improve the conditioning of the system matrix and to denoise. This dataset is generated by initially generating $300$ points uniformly at random in the $2$-D plane. Then using the Delaunay triangulation, we generate triangular lattices and remove a few edges to create a simplicial complex with two holes, 300 nodes, 783 edges, and 505 triangles as in Fig.~\ref{fig:fig2}.  We follow a similar procedure as before to generate bandlimited signals on this simplicial complex using $W_{0}= 50$, $W_{2}=50$, and $R_{1}= 2$.  We also add zero mean white Gaussian noise of variance $10^{-5}$ to $\vx_1$ and choose $P = 10$ and $|\cS|=50$. These $50$ observed edge signals are shown in Fig.~\ref{fig:fig2}.  We can observe in Figs.~\ref{fig:fig2}(b)-(d) that the reconstructed simplicial signals (bottom) are very close to the true simplicial signals, which demonstrates the efficacy of the developed method in recovering multi-order simplicial signals by subsampling a few edge signals and its robustness to noisy observations. Further, to test the robustness of the proposed algorithm for different noise levels, we generate noisy $\vx_1$ by varying noise variance and evaluate the proposed algorithm for a fixed $P = 10$ and varying the size of the sampling set. In particular, for each noise variance, we generate $100$ realizations of noisy $\vx_{1}$ and evaluate $\rm{MSE}$, which is computed as ${\rm MSE} = ({\rm MSE}(\vx_0) + {\rm MSE}(\vx_2) + {\rm MSE}(\vr_1))/3$, where 
$
{\rm MSE}(\vx) = \mathbb{E}[\|\vx - \vx_{{\rm ls}}\|^2_2]$
with $\vx_{\rm ls}$ being the least squares estimator of $\vx$. In Fig.~\ref{fig:fig2}(e), we report the mean squared error against noise variance for three different sizes of the sampling set. It can be observed the proposed algorithm recovers the multi-order signals perfectly in the noiseless case.
As expected, $\rm{MSE}$ increases with the increase in variance or decrease in the size of the sampling set.   

\section{Conclusions}
We proposed a method for recovering multi-order bandlimited simplicial signals from subsampled aggregated edge flow signals. The proposed algorithm leveraged the Helmholtz decomposition to relate simplicial signals of different orders.  For recovery, we proposed a simple least squares estimator. We also provided conditions on the number of aggregations and the size of the sampling set as a function of the bandwidth of the simplicial signals to be recovered for perfect recovery.  


\appendix
In this section, we relate the eigenvectors $\mQ_{0} \in  \rm{range}(\mL_0)$ and $\mQ_{2} \in \rm{range}(\mL_2)$ to eigenvectors $\mU_{\rm low} \in \rm{range}(\mL_{\rm low})$ and $\mU_{\rm up} \in \rm{range}(\mL_{\rm up})$, respectively, for self containment.
We have for $\mL_0 = \mB_1\mB_1^T$, $\mB_1\mB_1^T\mQ_0 = \mQ_0\bLambda_0$. On multiplying both sides with $\mB_1^T$, we get 
$\mB_1^T\mB_1\mB_1^T\mQ_0 = \mL_{\rm low}\mB_1^T\mQ_0 = \mB_1^T\mQ_0\bLambda_0$, which implies that eigenvectors of $\mL_{\rm low}$ are given by $\mU_{\rm low} = \mB_1^T\mQ_0$ as in \eqref{eq:eigenvector_relations}.  Along the similar lines, we also have  $\mB_2\mB_2^T\mB_2\mQ_2 = \mB_2\mQ_2\bLambda_2$, which implies that $\mU_{\rm up} = \mB_2\mQ_2$.

\pagebreak
\bibliographystyle{IEEEtran}
\IEEEtriggeratref{10}
\bibliography{refs.bib}

\end{document}